\pdfoutput=1

\documentclass[11pt,twoside,a4paper,cmspaper,final,collab]{cms-tdr}
\def\svnVersion{444261P}\def\cmsCernNoTag{CERN-EP-2017-272}\def\cmsCernDate{\today}\def\cmsMessage{Published in Physics Letters B as \href{http://dx.doi.org/10.1016/j.physletb.2018.01.049}{\doi{10.1016/j.physletb.2018.01.049.}}}

\begin{document}\cmsNoteHeader{B2G-16-025}

\hyphenation{had-ron-i-za-tion}
\hyphenation{cal-or-i-me-ter}
\hyphenation{de-vices}
\RCS$HeadURL: svn+ssh://svn.cern.ch/reps/tdr2/papers/B2G-16-025/trunk/B2G-16-025.tex $
\RCS$Id: B2G-16-025.tex 440452 2018-01-03 14:36:04Z kfjack $

\newcommand{\tstar}{\ensuremath{{\PQt{}^{*}}}\xspace}
\newcommand{\atstar}{\ensuremath{{\PAQt{}^{*}}}\xspace}
\newcommand{\tstarpair}{\ensuremath{\tstar\atstar}}

\newcommand{\xplusjet}[1]{\ensuremath{{#1}{+}\text{jet}}}
\newcommand{\xplusjets}[1]{\ensuremath{{#1}{+}\text{jets}}}
\newcommand{\muplusjets}[1]{\xplusjets{\PGm}}
\newcommand{\elplusjets}[1]{\xplusjets{\Pe}}
\newcommand{\topplusjet}{\xplusjet{\PQt}}
\newcommand{\threehalf}{\ensuremath{3\hspace{-.1em}/\hspace{-.1em}2}\xspace}
\newcommand{\onehalf}{\ensuremath{1\hspace{-.1em}/\hspace{-.1em}2}\xspace}
\newcommand{\enc}[1]{\ensuremath{\left( #1 \right)}}
\newcommand{\x}{\ensuremath{\phantom{0}}}
\newlength\cmsFigWidth\ifthenelse{\boolean{cms@external}}{\setlength\cmsFigWidth{0.95\columnwidth}}{\setlength\cmsFigWidth{0.45\textwidth}}
\newlength\cmsFigWidthLarge\ifthenelse{\boolean{cms@external}}{\setlength\cmsFigWidthLarge{0.95\columnwidth}}{\setlength\cmsFigWidthLarge{0.60\textwidth}}

\cmsNoteHeader{B2G-12-014}
\title{Search for pair production of excited top quarks in the lepton+jets final state}

\author{The CMS Collaboration}

\date{\today}

\abstract{A search is performed for the pair production of spin-\threehalf excited top quarks, each decaying to a top quark and a gluon. The search uses the data collected with the CMS detector from proton-proton collisions at a center-of-mass energy of 13\TeV, corresponding to an integrated luminosity of 35.9\fbinv. Events are selected by requiring an isolated muon or electron, an imbalance in the transverse momentum, and at least six jets of which exactly two must be compatible with originating from the fragmentation of a bottom quark. 
No significant excess over the standard model predictions is found.  
A lower limit of 1.2\TeV is set at 95\% confidence level on the mass of the spin-\threehalf excited top quark in an extension of the Randall--Sundrum model, assuming a 100\% branching fraction of its decay into a top quark and a gluon.
These are the best limits to date in a search for excited top quarks and the first at 13\TeV.
}

\hypersetup{%
pdfauthor={CMS Collaboration},%
pdftitle={Search for pair production of excited top quarks in the lepton+jets final state},%
pdfsubject={CMS},%
pdfkeywords={CMS, physics, beyond two generations, excited top quark}}

\maketitle

\section{Introduction}
\label{section:Introduction}

The standard model (SM) of particle physics provides a successful description of the properties of the elementary particles and their interactions.
Despite its success, the SM is assumed to be an effective model of a more complete theory.
Many extensions of the SM predict that the top quark is a composite particle and not a fundamental object~\cite{
comptop:PhysRevD.51.3888,
comptop:Lillie:2007hd,
comptop:PhysRevD.78.074026,
comptop:Kumar:2009vs}.
A direct confirmation of this hypothesis could be achieved by the discovery of an excited top quark (\tstar).

In models that describe the proposed excited top quark~\cite{excitedquarklep:PhysRevD.42.815,excitedquarklim:Harris:1996ct}, weak isodoublets are used to represent both left- and right-handed components of the \tstar~quark, allowing for a description of finite masses prior to the onset of electroweak symmetry breaking.
Thus, in contrast to the heavy top quark from a sequential fourth-generation model, in these models the existence of \tstar~quarks is not strongly constrained by the discovery of a SM-like Higgs boson~\cite{higgs:atlascms:Aad:2015zhl,higgscms:Chatrchyan:2012xdj,higgsatlas:Aad:2012tfa}.
In string realizations of the Randall--Sundrum~(RS) model~\cite{rs_scenario:PhysRevLett.83.4690,rs_scenario:PhysRevLett.83.3370},
the right-handed \tstar~quark is expected to be the lightest spin-\threehalf excited state~\cite{stringres:Hassanain:2009at}.

A spin-\threehalf \tstar~quark is described by the Rarita--Schwinger~\cite{spinlagrangian:PhysRev.60.61} vector spinor Lagrangian.
At the energy of LHC, the production cross section of spin-\threehalf quarks is proportional to~$\hat{s}^3$, where~$\hat{s}$ is the square of the energy in the parton-parton collision rest frame, rather than $\hat{s}^{-1}$, as it is for spin-\onehalf quarks~\cite{tstarxsec:Stirling:2011ya}.
Therefore, when integrating over the parton momentum fractions~($x$) in proton-proton collisions, 
spin-\threehalf quarks receive a contribution at large~$x$ values that is greater than that from spin-\onehalf quarks.
In the RS model, the spin-\threehalf \tstar~quark is expected to have a pair production cross section of the order of a few picobarns at $\sqrt{s}=13\TeV$, for a \tstar~of mass $m_{\tstar}=1\TeV$
~\cite{comptop:PhysRevD.51.3888,spin32quark:PhysRevD.87.015023,tstarxsec:Stirling:2011ya}, which dominates over single $\tstar$ production for most of the parameter space in the model~\cite{stringres:Hassanain:2009at}. 
The \tstar~quark decays predominantly to a top~quark through the emission of a
gluon~\cite{
stringres:Hassanain:2009at,
comptop:PhysRevD.51.3888,
spin32quark:PhysRevD.87.015023,
spin32:PhysRevD.39.1883}.

In this Letter, we present a search for pair-produced \tstar~quarks, where each \tstar~quark decays exclusively to a top quark (\PQt) and a gluon (\Pg).
We use data recorded in 2016 with the CMS detector in proton-proton ($\Pp\Pp$) collisions at $\sqrt{s}=13\TeV$ at the LHC, corresponding to an integrated luminosity of 35.9\fbinv.
We consider the case where one top quark decays via a hadronically decaying W boson, and the W boson originating from the second top quark decays to an electron or muon and a neutrino:
$\tstarpair \rightarrow (\PQt\Pg)(\PAQt\Pg) \rightarrow  (\PW\PQb\Pg)(\PW\PQb\Pg) \rightarrow (\PQq\PQq^\prime\PQb\Pg)(\ell\PGn\PQb\Pg)$.
We refer to the resulting final state (one reconstructed muon or electron, missing transverse momentum, and multiple jets) as the lepton+jets decay topology. 

A search for pair-produced \tstar~quarks was previously performed by CMS using $\Pp\Pp$~collisions at $\sqrt{s}=8\TeV$~\cite{tstar8tev:Chatrchyan:2013oba}.
This Letter presents a more sensitive search because of the higher collision energy and therefore larger signal cross sections, 
and the larger data sample, which is nearly twice the size.
In addition, the simulation has been improved by explicitly including the Rarita--Schwinger Lagrangian in the generator, resulting in the correct spin correlations for the signal.

\section{The CMS detector and simulated samples}
\label{sec:data}

The central feature of the CMS apparatus is a superconducting solenoid of 6\unit{m} internal diameter, providing a magnetic field of 3.8\unit{T}.
Within the solenoid volume are a silicon pixel and strip tracker, a lead tungstate crystal electromagnetic calorimeter (ECAL), and a brass and scintillator hadron calorimeter (HCAL), each composed of a barrel and two endcap sections.
Forward calorimeters extend the pseudorapidity ($\eta$) coverage provided by the barrel and endcap detectors.
Muons are measured in gas-ionization detectors embedded in the steel flux-return yoke outside the solenoid.
A more detailed description of the CMS detector, together with a definition of the coordinate system used and the relevant kinematic variables, can be found in Ref.~\cite{cmsdectector:Chatrchyan:2008aa}.

Simulated $\tstarpair$ signal events are generated in 100\GeV steps with $m_{\tstar}$ in the range 700--1600\GeV, using the \MGvATNLO~\cite{madgraphnlo:Alwall:2014hca} event generator and NNPDF3.0~\cite{nnpdf:Ball:2014uwa} for the parton distribution functions (PDFs).
The \tstarpair~production cross section ranges from ${\approx}5\unit{pb}$ at $m_{\tstar} = 700\GeV$, down to ${\approx}4\unit{fb}$ at $m_{\tstar} = 1600\GeV$.
This cross section is calculated at leading order in perturbation theory, with the factorization and renormalization scales set to $m_{\tstar}$;
the calculation is cut off at $7m_{\tstar}$ to prevent unitarity violation.
The Rarita--Schwinger Lagrangian, included in the \MADGRAPH~5 generator, is used for simulating spin-\threehalf \tstarpair~events.
This implementation and the corresponding physics parameters are provided by the authors of Ref.~\cite{tstarxsec:Stirling:2011ya}.
The width of the ${\tstar}$ quark is assumed to be 10\GeV, which is much narrower than the detector resolution.
Parton shower and hadronization processes are modeled using \PYTHIA~8.212~\cite{pythia8:Sjostrand2015159}.
The generated events are processed through a simulation of the CMS detector based on \GEANTfour~\cite{geant4:Agostinelli2003250}, and are reconstructed using the same algorithms as used for data.

We estimate SM backgrounds using a data-derived approach.
Simulated samples for SM processes are used to study the modeling of the background and to provide a cross-check of the analysis procedures.
The simulated SM samples relevant to this analysis are: \ttbar~production; single top quark production via the $s$-channel, $t$-channel, and tW processes; $\PW$~and $\PZ$~boson production in association with jets; the $\ttbar{+}\PW$, $\ttbar{+}\PH$, and $\ttbar{+}\PZ$ processes.
The \ttbar~and $\ttbar{+}\PH$ processes are simulated using \POWHEG 2.0~\cite{
powheg:Nason:2004rx,
powheg:Frixione:2007vw,
powheg:Alioli:2010xd,
powheg:ttbar:Campbell:2014kua,
powheg:ttH:Hartanto:2015uka},
while the other SM processes are simulated using
\MGvATNLO up to next-to-leading order~\cite{
madgraphnlo:Alwall:2014hca,
madgraphFXFXmix:Frederix:2012ps,
madgraphMLMmix:Alwall:2007fs}.
All simulated samples include the additional contributions from overlapping $\Pp\Pp$~collisions within the same and nearby bunch crossings~(``pileup'') at large instantaneous luminosity.
Simulated events are given individual weights to match the distribution of the average number of pileup interactions in data.

\section{Event reconstruction}
\label{sec:reconstruction}

Event reconstruction is based on the CMS particle-flow (PF) algorithm~\cite{pfalgo:CMS-PRF-14-001}, which takes into account information from all subdetectors, including measurements from the tracking system, energy deposits in the ECAL and HCAL, and tracks reconstructed in the muon detectors. Given this information, all particles in the event are reconstructed as electrons, muons, photons, and charged or neutral hadrons.
Photons
are identified as ECAL energy clusters not linked to the extrapolation of any charged-particle trajectory to the ECAL.
Muons
are identified as a track in the central tracker consistent with either a track or several hits in the muon system, 
and not associated with energy clusters in the calorimeters.
Electrons
are identified as 
a primary charged particle track that extrapolates to at least one ECAL energy cluster.  The track may be associated with bremsstrahlung photons emitted
along the way through the tracker material.
Charged hadrons are identified as charged-particle tracks neither identified as electrons, nor as muons.
Finally, neutral hadrons are identified as HCAL energy clusters not linked to any charged-hadron trajectory, or to ECAL and HCAL energy excesses with respect to the expected charged hadron energy deposits.

For each event, jets from these reconstructed particles are clustered with the infrared and collinear safe anti-\kt algorithm~\cite{antikt:Cacciari:2008gp}, using a distance parameter $R=0.4$. Charged hadrons associated with pileup vertices are excluded from jet reconstruction.
The jet momentum is the vectorial sum of the momenta of all particles contained in the jet.
The reconstructed jet momentum is found in simulation to be within 5 to 10\% of the true momentum over the whole \pt spectrum and detector acceptance.
Jet energy corrections are derived from the simulation and measurements in collision data~\cite{jetres:Khachatryan:2016kdb}.
The jet energy resolution amounts typically to 15\% at 10\GeV, 8\% at 100\GeV, and 4\% at 1\TeV~\cite{jetres:Khachatryan:2016kdb}.
The jet energy resolution in simulation is degraded to match that observed in data.

Jets are identified as originating from a bottom quark through a combined secondary vertex algorithm~CSVv2~\cite{Chatrchyan:2012jua, btag:CMS-PAS-BTV-15-001}.
The algorithm uses a multivariate discriminator to combine information on the significance of the impact parameter, the jet kinematics, and the location of the secondary vertex.
A working point of the discriminator with ${\approx}70\%$ \PQb~quark identification efficiency and ${\approx}1\%$ mistag efficiency for light quarks and gluons is used in this analysis.
Small differences in $\PQb$~tagging efficiencies and mistag rates between data and simulated events are accounted for by applying additional corrections to simulation.

The missing transverse momentum vector is defined as the negative vector sum of the momenta of 
all reconstructed PF candidates in an event projected onto the plane perpendicular to the beams.
Its magnitude is referred to as \ptmiss.

\section{Event selection}
\label{sec:selection}

This analysis searches for \tstarpair~production, with each \tstar~decaying to $\PQt{+}\Pg$ and the \ttbar~pair in the event reconstructed in the lepton+jets final state. Events are required to contain exactly one isolated lepton, \ptmiss, and at least six jets, exactly two of which must be $\PQb$ tagged.

Events containing a muon are selected with a single-muon trigger that requires the presence of an isolated muon with transverse momentum $\PT>27\GeV$.
Events containing an electron are selected with a single-electron trigger that requires the presence of an isolated electron with $\PT>32\GeV$.
The background rate for the single electron trigger was much higher than for the single muon trigger, requiring more stringent selection criteria for the electron channel.
A deterministic annealing algorithm is used to reconstruct the candidate primary vertices~\cite{Chatrchyan:2014fea}; the vertex with the highest track multiplicity is selected as the primary event vertex. Selected events are required to have this primary vertex within 2\cm of the center of the detector in the $x$-$y$ plane, and within 24\cm along the $z$-direction.

Offline, muons are required to have $\PT>30\GeV$ and $\abs{\eta}<2.1$.
The track associated with a muon is required to have hits in the pixel and muon detectors, a good quality fit, and transverse and longitudinal impact parameters with respect to the primary vertex smaller than 2 and 5\mm, respectively.
An isolation factor $I$ is defined as the scalar sum, divided by the muon \PT, of the \PT of all
photons, charged hadrons, and neutral hadrons within an angular cone of $\Delta R \equiv \sqrt{\smash[b]{(\Delta \eta)^2 + (\Delta \phi)^2}} < 0.4$ (where $\phi$ is the azimuthal angle) around the track, corrected for the effects of pileup~\cite{muoniso:Chatrchyan:2012xi}.
An isolation selection $I<0.15$, corresponding to an efficiency of ${\approx}95\%$ is used.

Electrons are required to have $\PT>35\GeV$ and to be within the region $\abs{\eta}<2.1$.
Electrons within $1.44 < \abs{\eta} < 1.56$, corresponding to the ECAL barrel--endcap transition region, are rejected to avoid poor reconstruction performance.
Electrons are selected using a cutoff-based selection method~\cite{Khachatryan:2015hwa}
based on the shower shape, the track quality, the spatial match between the track and the electromagnetic cluster, the fraction of total cluster energy in the HCAL, and the resulting level of activity in the surrounding tracker and calorimeter regions.
The criteria imposed in these electron selection algorithms have a combined efficiency of ${\approx}70\%$.

In addition to the selections above, the leptons are required to have an angular separation $\Delta R<0.1$ with respect to the lepton reconstructed by the trigger system.
The lepton selection efficiencies for data and simulation are measured using the 
tag-and-probe method~\cite{Khachatryan:2015hwa}.
Additional corrections are applied to simulation to account for observed differences in the efficiencies between data and simulation.

The \ptmiss is required to be greater than $20\GeV$, while the jets are required to have $\PT>30\GeV$, $\abs{\eta}<2.4$, and angular separation
$\Delta R>0.4$ with respect to well-identified electrons or muons.
In order to reject misreconstructed, poorly reconstructed, and noisy jets,
the fractional energy contribution from both ECAL and HCAL must be non-zero and non-unity.
Exactly two jets are required to pass the \PQb~tagging criteria.

The expected yields after event selection are summarized in Table~\ref{tab:selection_summary}.
Simulated signal events pass the selection criteria with acceptance times efficiency of 1.4--2.2\%, depending on the channel and on the signal mass.
After the application of all selections, $44\,573$ events are observed in the $\mu$+jets~ channel and $28\,942$ events in the e+jets~ channel.
The yields predicted from the simulated SM background processes are $46\,600$ events in the $\mu$+jets~ channel and $30\,700$ events in the e+jets~ channel.

\begin{figure*}
\centering
\includegraphics[width=\cmsFigWidth]{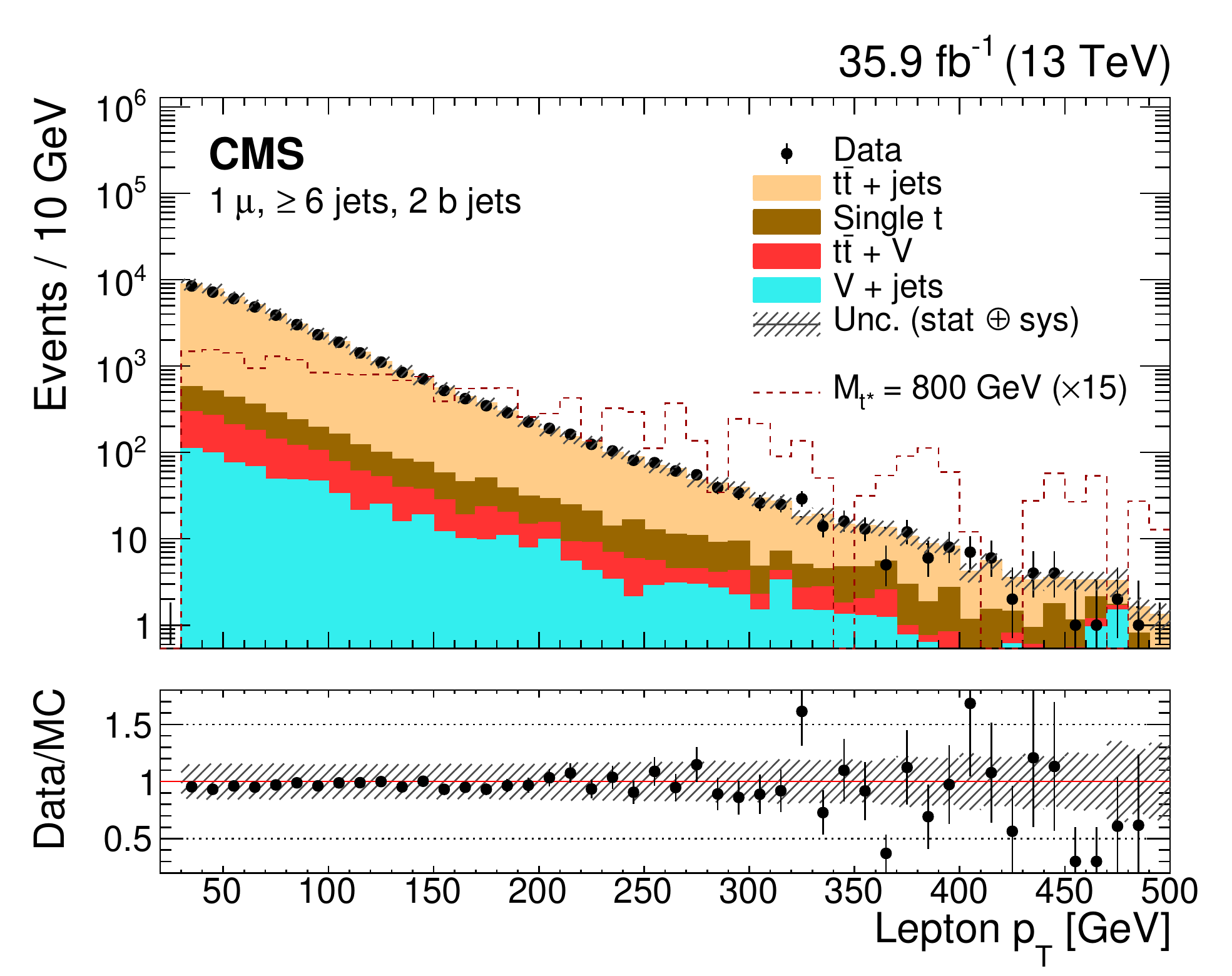} \hfil
\includegraphics[width=\cmsFigWidth]{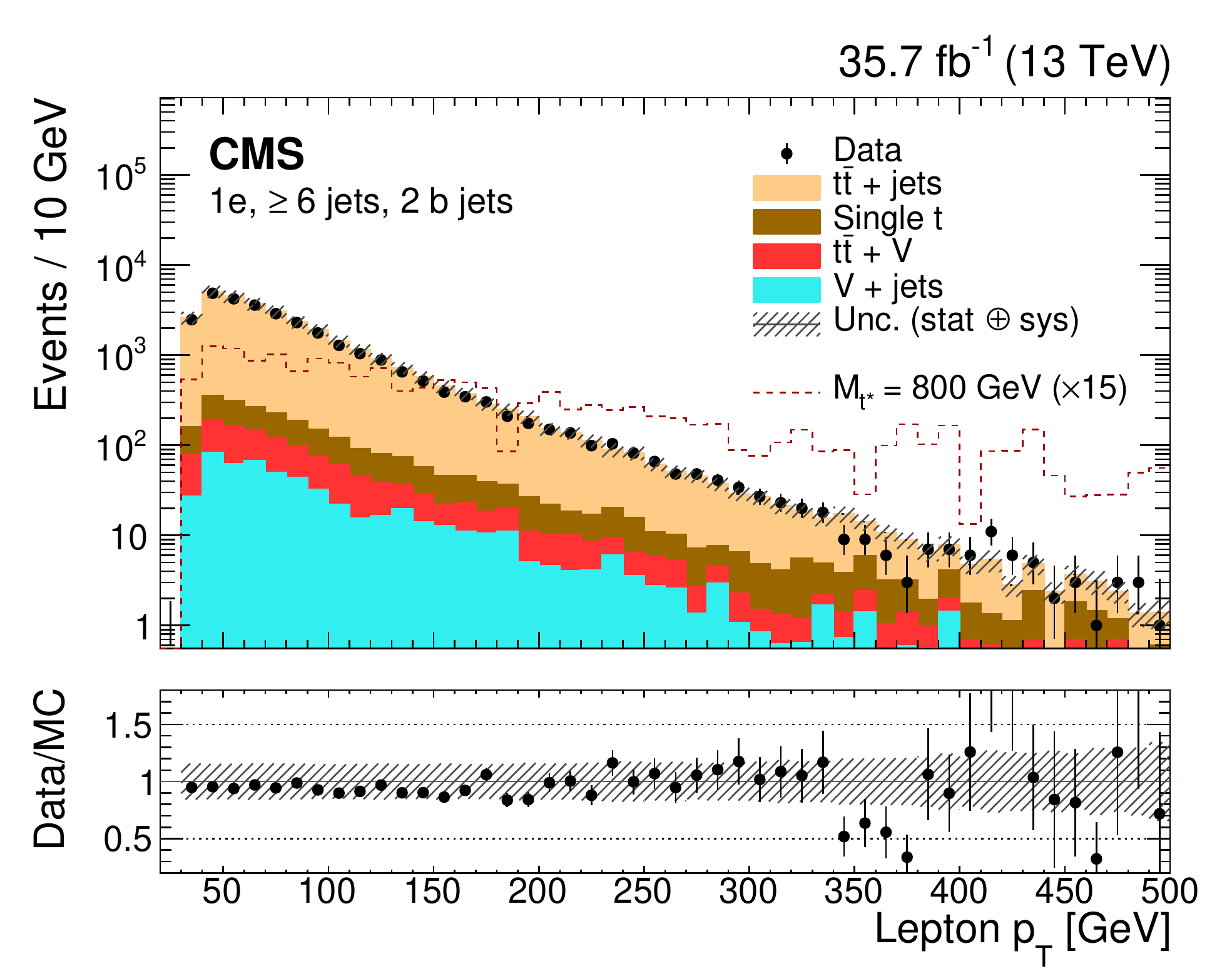} \\
\includegraphics[width=\cmsFigWidth]{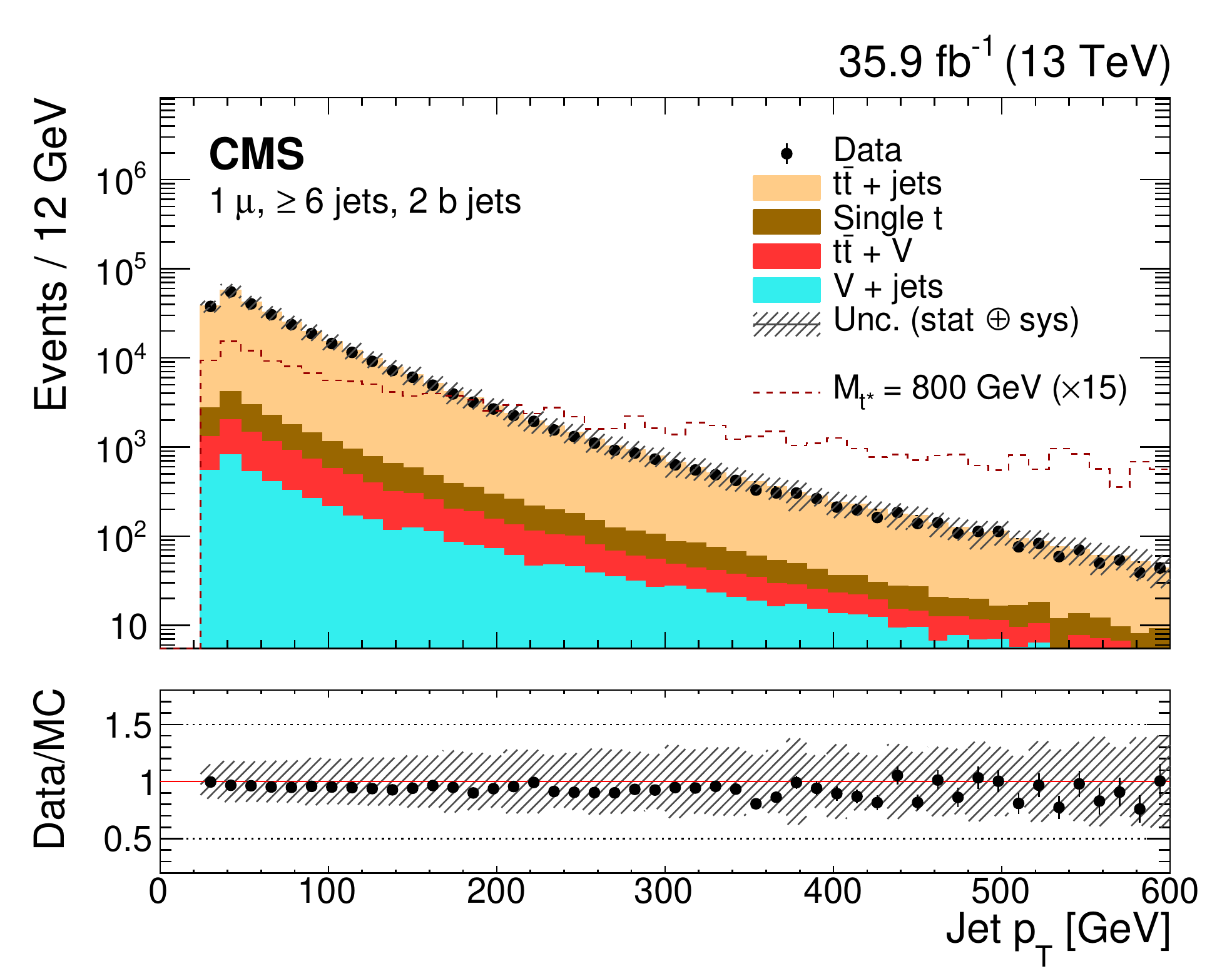} \hfil
\includegraphics[width=\cmsFigWidth]{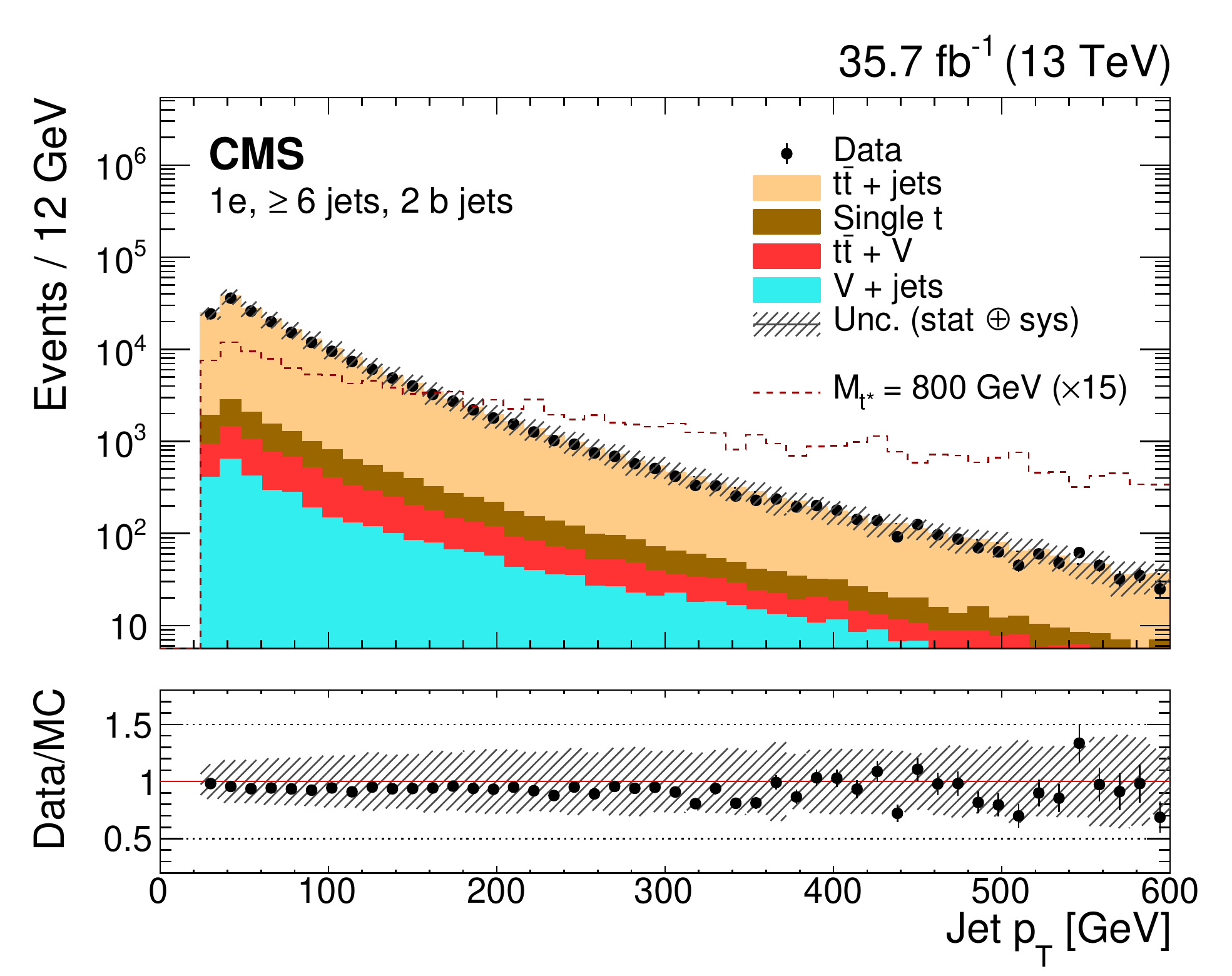} \\
\includegraphics[width=\cmsFigWidth]{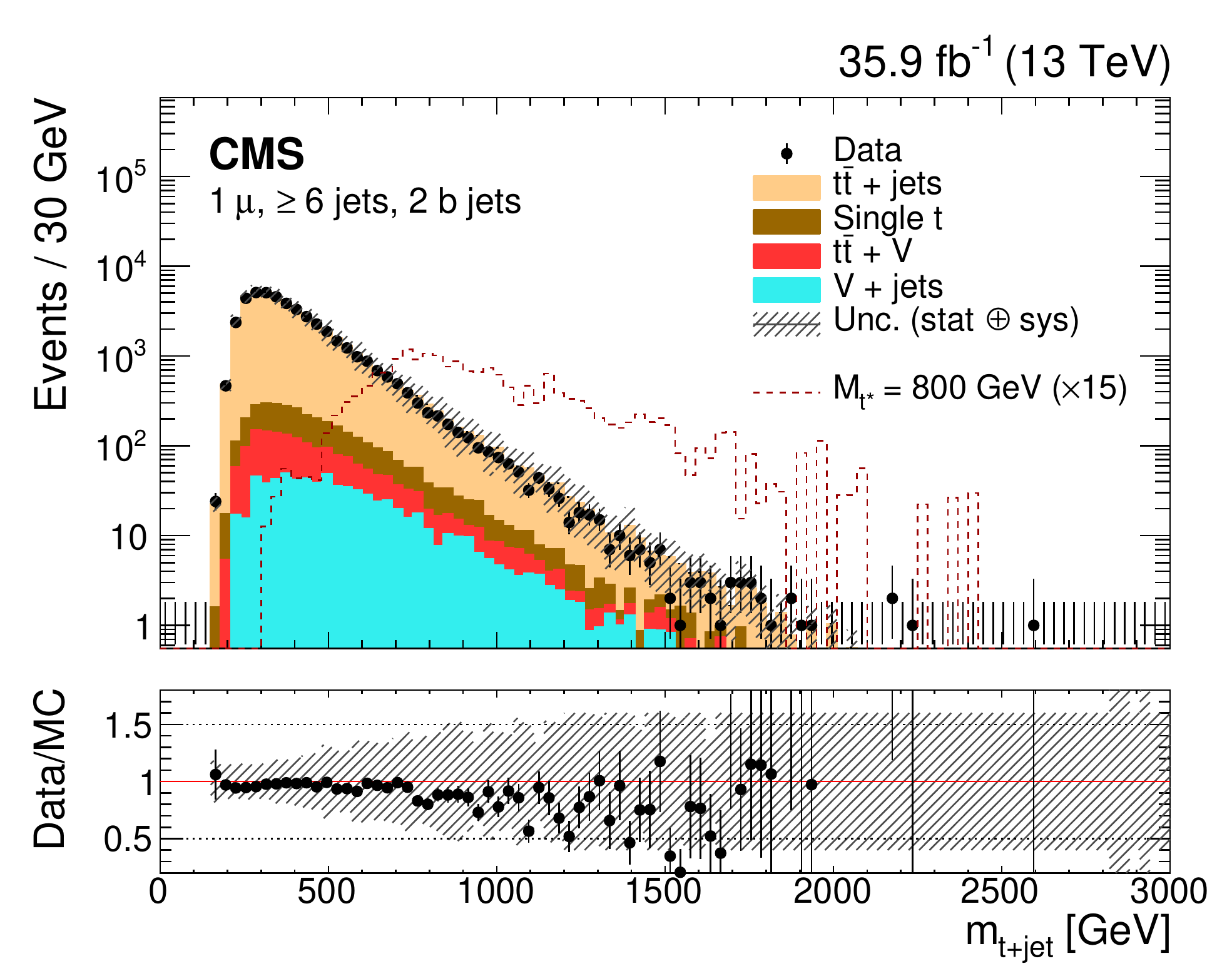} \hfil
\includegraphics[width=\cmsFigWidth]{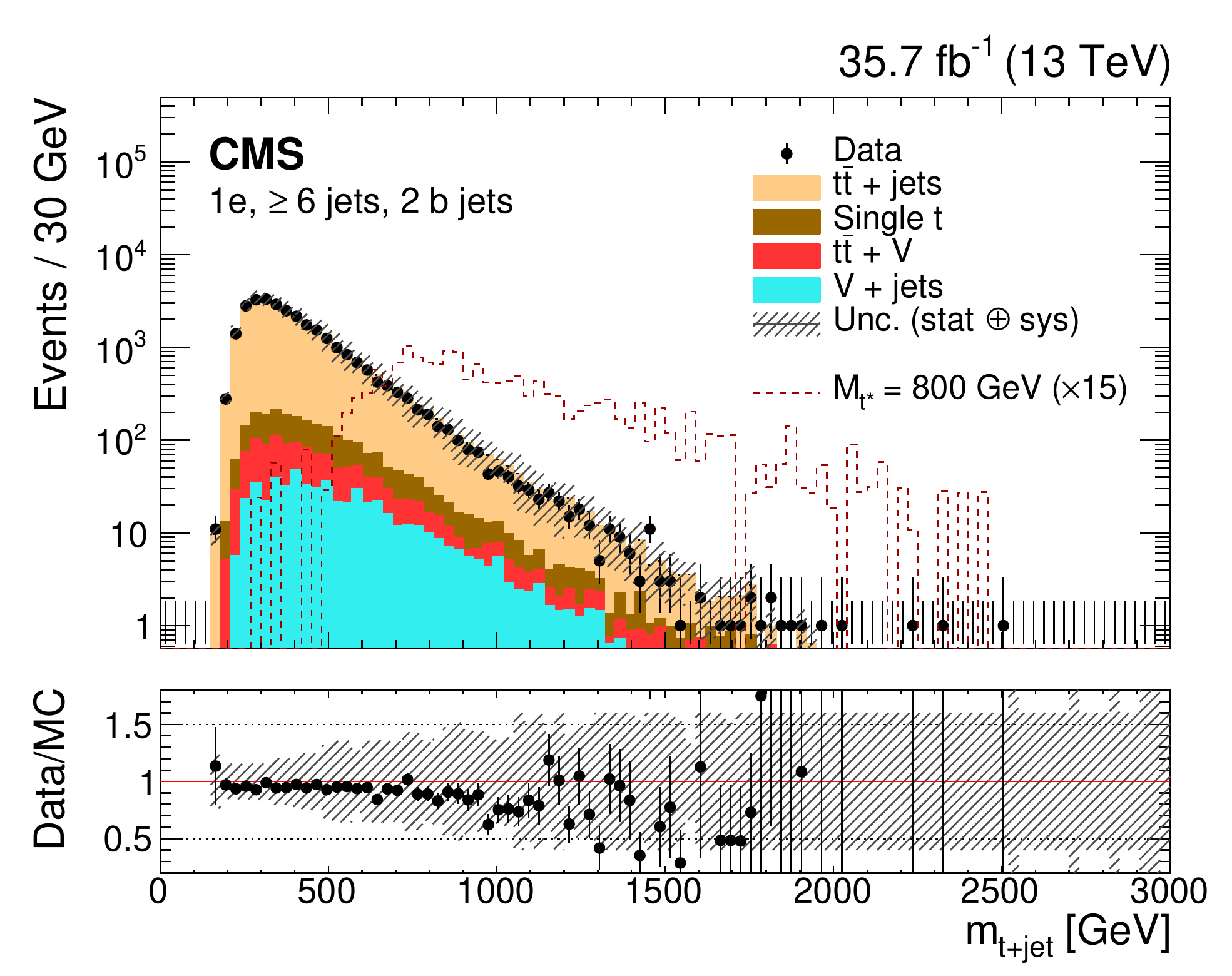}
\caption{
Kinematic distributions of selected events with a single lepton and six or more jets of which exactly two are $\PQb$ tagged. 
Data events (points), simulated background processes (stacked histograms), and a simulated 800\GeV signal process (dashed line) are shown.
Events selected in the $\PGm$+jet final state are shown on the left while those in the $\Pe$+jet final state are shown on the right.
From upper to lower, the kinematic variables displayed are the lepton \PT, the jet \PT and the  $m_{\rm t+jet}$.
The shaded region is the total uncertainty of the simulated background processes, which includes statistical and systematic uncertainties.
}
\label{fig:kincomp}
\end{figure*}

Small differences between data and the SM predictions are within the estimated uncertainties of the simulation, with the dominant uncertainty being the choice of the renormalization and factorization scales used in the generator of the \ttbar~events.
Details of the uncertainties are given in Section \ref{sec:uncertainties}.
Furthermore, the differential distributions of kinematic variables of simulated SM processes are also in agreement with data, as shown in Fig.~\ref{fig:kincomp}.
In particular, the distribution of the invariant mass of a $\PQt$+jet~system ($m_{\rm t+jet}$, see Section \ref{sec:mass} for details) in data is in agreement with the background estimation.

\begin{table}
\topcaption{
Expected numbers of selected events for the simulated signal process
as a function of $ m_{\tstar}$. Also shown are the expected numbers of events predicted by the SM, together with the systematic uncertainties discussed in Section~\ref{sec:uncertainties} and the uncertainties in the cross sections of the various processes, as well as the numbers of selected events observed in data. }
\label{tab:selection_summary}
\centering
\ifthenelse{\boolean{cms@external}}{\resizebox{\columnwidth}{!}}{}
{
\begin{tabular}{cccc}
\hline
& $\PGm$+jet final state & $\Pe$+jet final state \\
\hline
{$\tstarpair$ signal, $ m_{\tstar}$} &  & \\
$\x700\GeV$  & 3670 & 2730 \\
$\x800\GeV$  & 1230 & 1010 \\    
$\x900\GeV$  & 483  & 369  \\
$1000\GeV$ & 200  & 148  \\
$1100\GeV$ & 92   & 69   \\
$1200\GeV$ & 40   & 29   \\
$1300\GeV$ & 20   & 15   \\
$1400\GeV$ & 9    & 7    \\
$1500\GeV$ & 4    & 4    \\
$1600\GeV$ & 2    & 2    \\
\hline
SM processes & $(4.66\pm0.38){\times}10^{4}$ & $(3.07\pm0.23){\times}10^{4}$\\
\hline
Data         & 44\,573                     & 28\,942 \\
\hline
\end{tabular}
}
\end{table}

\section{Mass reconstruction}
\label{sec:mass}
Since the dominant background is SM \ttbar~production with extra jets, the reconstructed invariant mass spectrum of the $\PQt$+jet~systems is used to distinguish between \tstarpair~signal and \ttbar~background.
The \ptmiss is assumed to be carried away entirely by the neutrino from the leptonically decaying $\PW$~boson ($\PW_{\text{lep}}$). 
We assume that the parent $\PW$~boson is on shell and the neutrino is massless in order to determine the longitudinal momentum of the neutrino.

Given the high jet multiplicity of the event selection, a measure was designed for evaluating different associations of the reconstructed jets with the parton objects in the final state.
For the jets, the six jets with the highest $\PT$ values are taken into consideration. 
The $\PQb$ tagged jets are assigned to one of the $\PQb$ quark partons, and the other jets are associated 
with the decay daughters of the hadronically decaying $\PW$ ($\PW_{\text{had}}$) or with the gluons from $\tstar$ decay. 
The quality of the jet-parton assignment for a single event is evaluated with an $S$ value based on how well the intermediate physical objects are reconstructed:
\begin{equation}
\ifthenelse{\boolean{cms@external}}
{
\begin{split}
S = & \enc{ \frac{m_{\PQq\PQq'} - m_{\PW}}{\sigma_{\PW}} }^2
+ \enc{ \frac{m_{\PQq\PQq'\PQb} - m_{\PQt}}{\sigma_{\PQt,\text{had}}}}^2 \\ 
 + &  \enc{ \frac{m_{\ell\PGnl\PQb} - m_{\PQt}}{\sigma_{\PQt,\text{lep}}}}^2
+ \enc{ \frac{m_{\PQq\PQq'\PQb\Pg} - m_{\ell\PGnl\PQb\Pg}}{\sigma_{\tstar}}}^2,
\end{split}
}
{
S = \enc{ \frac{m_{\PQq\PQq'} - m_{\PW}}{\sigma_{\PW}} }^2
+ \enc{ \frac{m_{\PQq\PQq'\PQb} - m_{\PQt}}{\sigma_{\PQt,\text{had}}}}^2
+ \enc{ \frac{m_{\ell\PGnl\PQb} - m_{\PQt}}{\sigma_{\PQt,\text{lep}}}}^2
+ \enc{ \frac{m_{\PQq\PQq'\PQb\Pg} - m_{\ell\PGnl\PQb\Pg}}{\sigma_{\tstar}}}^2,
}
\end{equation}
where $m_{\PQq\PQq'}$ is the invariant mass of the jets assigned to $\PW_{\text{had}}$ daughters.
Invariant masses of the physical objects assigned to hadronically 
and leptonically decaying $\PQt$ ($\tstar$) quarks
are denoted by $m_{\PQq\PQq'\PQb}$ ($m_{\PQq\PQq'\PQb\Pg}$) and 
$m_{\ell\PGnl\PQb}$ ($m_{\PQq\PQq'\PQb\Pg}$), respectively. 
$m_{\PW}$ and $m_{\PQt}$ are the mass of the $\PW$~boson and top quark recorded by the particle data group~\cite{pdg2016}, being 80.4 and 173.34\GeV, respectively.
The expected detector resolutions of the intermediate particles $\sigma_{\PW}$, $\sigma_{\PQt,\text{had}}$, $\sigma_{\PQt,\text{lep}}$ and $\sigma_{\tstar}$ are
estimated to be 24, 34, 30, and 230\GeV, respectively.
These estimates are obtained by reconstructing the $\tstarpair$, $\ttbar$~ and $\PW_{\text{had}}$ in the decay topology using the truth information from simulated signal samples.
Additional studies have shown that the mass reconstruction is insensitive to changes in the detector resolution values.

The jet-parton assignment with the smallest $S$ value is taken to represent the decay topology of a single event, under the \tstar~hypothesis.
The average value of the $m_{\PQq\PQq'\PQb\Pg}$ and $m_{\ell\PGnl\PQb\Pg}$ computed for this assignment is taken to represent the reconstructed \tstar~mass of an event, notated as $m_{\rm t+jet}$.
The rate at which all six jets are all correctly assigned is around 11\%, 
with the main difficulty being the correct assignment of the jets from the hadronically decaying $\PW$.

\section{Background modeling}
\label{sec:background}

To determine the presence of signal events in data, an unbinned extended maximum likelihood fit of a signal-plus-background model is performed on the $m_{\rm t+jet}>400\GeV$ spectrum.

The mass template of the \tstarpair~signal is constructed by smoothing the mass distribution from simulations, using an adaptive kernel estimation~\cite{kernelapprox:CRANMER2001198} with a Gaussian kernel and with no restriction on the boundary.
The smoothness parameter $\rho$ introduced in Ref.~\cite{kernelapprox:CRANMER2001198} is determined by the square root of the standard deviation of the signal distribution over the subset with ${\geq}4$ correctly assigned partons.

The background distribution is modeled using a log-normal function (up to a normalization factor):
\begin{equation}
f_{\text{bkg}}(m) = \frac{1}{m \sqrt{2\pi}} \exp\enc{ -a_2\ln^2\enc{\frac{m}{m_0}}},
\end{equation}
where $m$ is the mass, and $a_2$ and $m_0$ are the parameters that determine the shape of the background.
During the fit to the observed data, the number of background events, as well as the shape parameters of the background function, 
are free parameters.

To verify whether the fit is sensitive to the presence of \tstarpair~signal, a pseudo-data set is generated with the $m_{\rm t+jet}$ spectrum of the simulated backgrounds and then injected with the expected $m_{\rm t+jet}$ signal spectrum for various hypotheses of the signal cross section.
Performing the same fit over multiple sets of pseudo-data with varying signal cross sections showed no evidence of bias.

To ensure that the log-normal function is sufficient to model the background, a likelihood ratio test is conducted by comparing the results of fitting the spectrum of the simulated SM background to an extended log-normal functions of the form:
\begin{equation}\label{eqn:extlognorm}
\ifthenelse{\boolean{cms@external}}
{
\begin{split}
f_{\text{bkg},N}(m) = & \frac{1}{m \sqrt{2\pi}} \exp \left( 
-a_2\ln^2\enc{\frac{m}{m_0}}  \right. \\
 & \left. -a_3\ln^3\enc{\frac{m}{m_0}} 
 -\ldots
-a_N\ln^N\enc{\frac{m}{m_0}}
\right).
\end{split}
}
{
f_{\text{bkg},N}(m) = \frac{1}{m \sqrt{2\pi}} \exp\enc{
-a_2\ln^2\enc{\frac{m}{m_0}}
-a_3\ln^3\enc{\frac{m}{m_0}}
-\ldots
-a_N\ln^N\enc{\frac{m}{m_0}}
}.
}
\end{equation}
Increasing the number of parameters does not improve the description of the background.

\begin{figure*}
\centering
\includegraphics[width=\cmsFigWidth]{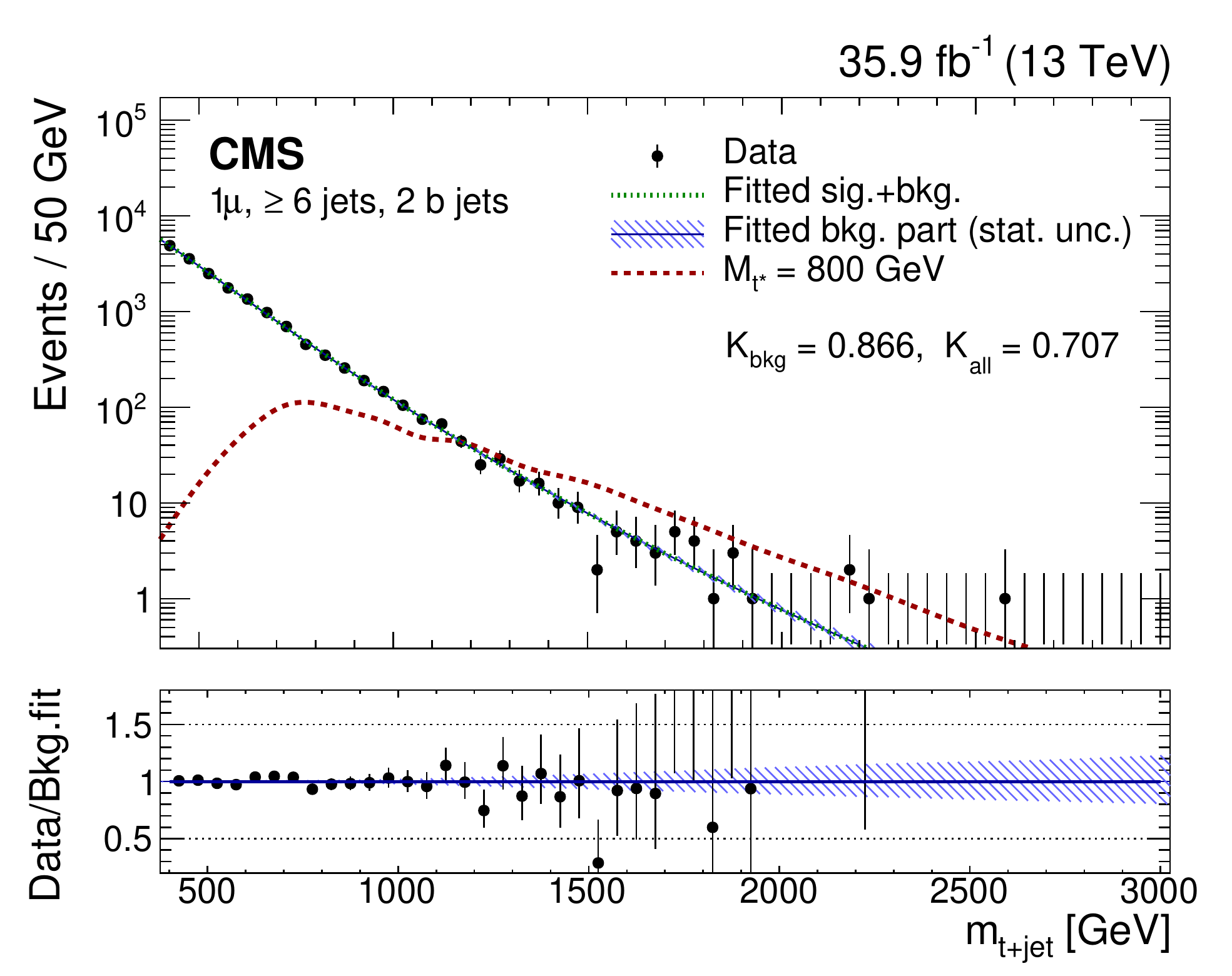} \hfil
\includegraphics[width=\cmsFigWidth]{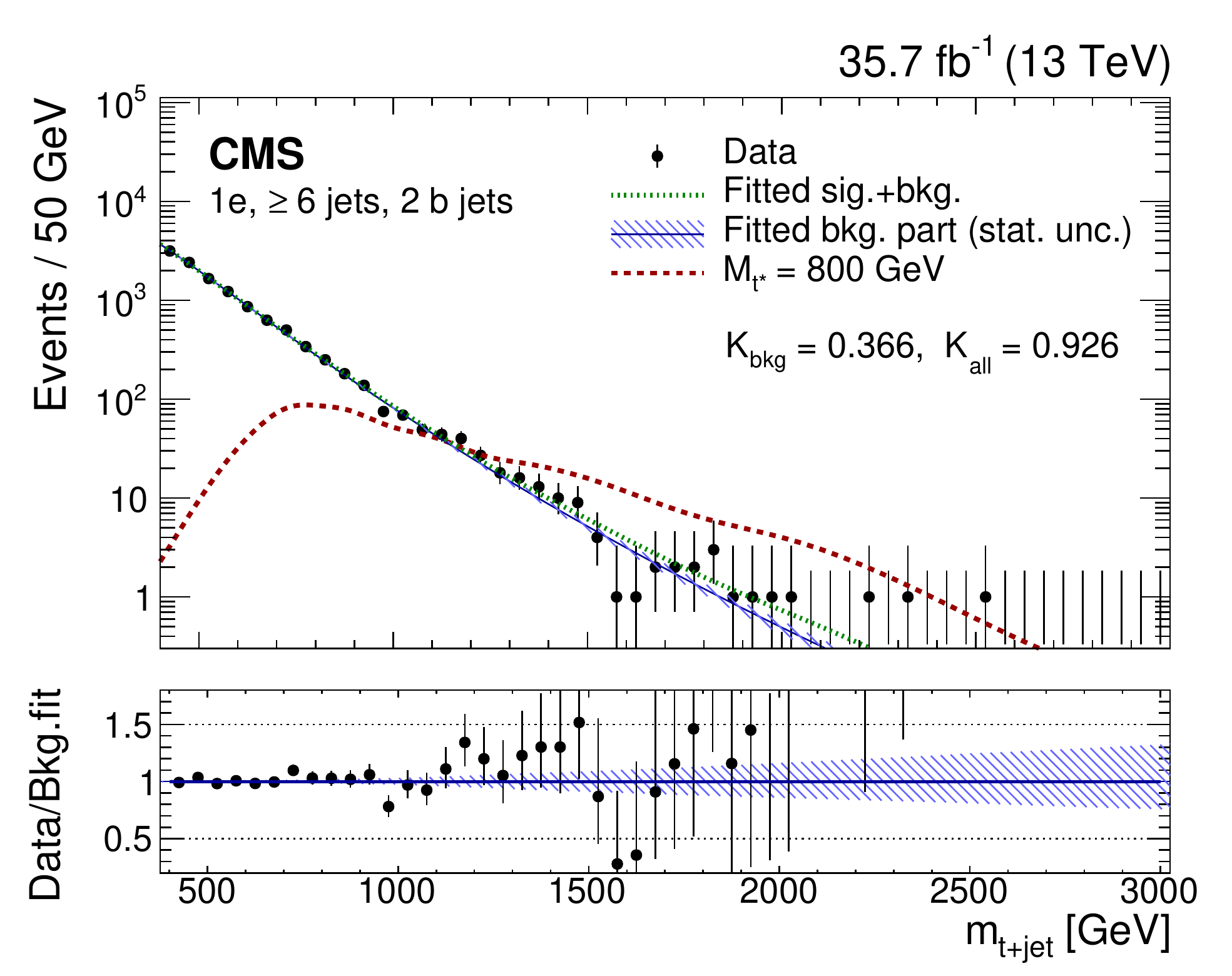}
\caption{
The $m_{\rm t+jet}$ spectrum for data (points), the signal+background fit (green), the background component of the signal+background fit (blue), and the expected spectrum for a simulated 800\GeV signal process (red dashed) normalized to the integrated luminosity of data.
Since there is no significant excess of signal found in data, the signal+background curve overlaps the background-only component.
The distributions for the $\PGm$+jets data are shown on the left 
while those for $\Pe$+jets data are shown on the right.
The probabilities of the Kolmogorov--Smirnov test between the data versus the signal+background model and between the data versus the background component are denoted by $K_{\text{all}}$ and $K_{\text{bkg}}$, respectively.
}
\label{fig:fitdemo}
\end{figure*}

The results of the fit performed on data with the 800\GeV signal spectrum are shown in Fig.~\ref{fig:fitdemo}.
The distribution of events in data is in agreement with a null hypothesis.
Based on the results of the Kolmogorov--Smirnov tests, the signal+background model and the background-only model both yield good fits to the data.

\section{Systematic uncertainties}
\label{sec:uncertainties}

The impact of experimental and theoretical sources of uncertainties is considered and summarized in Table~\ref{tab:systematics}.
For each source of uncertainty, alternative templates for the distribution of $m_{\rm t+jet}$ are generated by adjusting the relevant parameters in the simulation.

The uncertainties in the jet energy scale and jet resolutions depend on the $\PT$ and $\eta$ of the jets.
Alternative mass templates are generated by rescaling the nominal jet four-momentum in the simulation by ${\pm}1$ standard deviation (s.d.) of the associated uncertainties in energy scale and resolution.
Such uncertainties are also coherently propagated to all observables, including \ptmiss.
Varying the jet energy used for reconstruction has ${<}0.1\%$ impact on the signal acceptance.

The \PQb~tagging and lepton selection scale factors for residual differences between data and simulation have their respective systematic and statistical uncertainties.
Alternative templates are generated by shifting the correction scale factors by ${\pm}1$ s.d. for their respective uncertainties.
On average, the \PQb~tagging scale factor and lepton scale factors affect the signal acceptance by 2.8 and 2.5\%, respectively.

Because of uncertainties in the total inelastic $\Pp\Pp$~cross section, when calculating the data pileup scenario alternative pileup corrections are made with the inelastic cross section scaled by ${\pm}1$ s.d.
Variations in the pileup corrections have an average impact on the signal acceptance of 0.7\%.
The number of signal events is also affected by the uncertainty on the integrated luminosity, which is known to a precision of 2.5\%~\cite{CMS:2017sdi}.

The theoretical uncertainties considered are those associated with
the choice of the PDF, and the renormalization and factorization scales used by the event generator. 
The effects of the theoretical uncertainties are obtained by changing the various generator parameters within their estimated uncertainties and generating new $m_{\rm t+jet}$ fit templates that are used to calculate new sensitivities.

In addition to the statistical uncertainty originating from the signal+background fit, systematic uncertainties are introduced to cover the choice of modeling.
Alternative signal templates are generated with different choices of $\rho$ by changing the subset to require ${\geq}3$ and ${\geq}5$ correctly assigned partons.
The background shape is determined from data. Simulated events with different configurations, as well as several alternative models have been tested. The chosen model, with the parameters floated in the limit computation, has proven to describe the data and cover the associated systematic uncertainties sufficiently well.

\begin{table}
\centering
\topcaption{Sources of systematic uncertainties and the methods used to evaluate their effect on the simulated signal sample.
(s.d.: standard deviation, SF: correction scale factor)}
\label{tab:systematics}
\ifthenelse{\boolean{cms@external}}{
\resizebox{\columnwidth}{!}
{
\begin{tabular}{cc}
\hline
Source of uncertainty & \begin{tabular}{c} Implementation \\ on simulated signal sample  \end{tabular}     \\
\hline
Integrated luminosity          & \begin{tabular}{c} Normalization shift \\  by $\pm2.5\%$  \end{tabular}       \\
Statistical uncertainty        & \begin{tabular}{c} Normalization shift \\  by $\pm1$ s.d.        \end{tabular}        \\
Jet correction                 & \begin{tabular}{c} Correction factor varied  \\ by $\pm1$ s.d.   \end{tabular}       \\
Jet resolution                 & \begin{tabular}{c} Jet resolution shift \\  by $\pm1$ s.d.            \end{tabular}       \\
\PQb~tagging SF              & \begin{tabular}{c} SF varied \\ by $\pm1$ s.d.   \end{tabular}                      \\
Lepton efficiency SF         & \begin{tabular}{c} SF varied \\ by $\pm1$ s.d.   \end{tabular}                      \\
Pileup                         & \begin{tabular}{c} $\Pp\Pp$~inelastic cross section shifted \\ by $\pm4.6\%$~\cite{Aaboud:2016mmw}   \end{tabular}   \\
Modeling                       & \begin{tabular}{c} Smoothing parameter $\rho$ varied \\ over range $[1.17,1.66]$ \end{tabular} \\
PDF uncertainty                & \begin{tabular}{c} Generator parameter  varied \\ by $\pm1$ s.d.    \end{tabular}     \\
Scale uncertainty              & \begin{tabular}{c} Generator parameter varied \\  by $\pm1$ s.d.     \end{tabular}    \\
\hline
\end{tabular}
}
}
{
\begin{tabular}{cc}
\hline
Source of uncertainty & Implementation on simulated signal sample       \\
\hline
Integrated luminosity          & Normalization shift by $\pm2.5\%$                \\
Statistical uncertainty        & Normalization shift by $\pm1$ s.d.                \\
Jet correction                 & Correction factor varied  by $\pm1$ s.d.          \\
Jet resolution                 & Jet resolution shift by $\pm1$ s.d.                   \\
\PQb~tagging SF              & SF varied by $\pm1$ s.d.                        \\
Lepton efficiency SF         & SF varied by $\pm1$ s.d.                        \\
Pileup                         & $\Pp\Pp$~inelastic cross section shifted by $\pm4.6\%$~\cite{Aaboud:2016mmw}      \\
Modeling                       & Smoothing parameter $\rho$ varied over range $[1.17,1.66]$ \\
PDF uncertainty                & Generator parameter varied by $\pm1$ s.d.         \\
Scale uncertainty              & Generator parameter varied by $\pm1$ s.d.         \\
\hline
\end{tabular}
}
\end{table}

\section {Statistical analysis and extraction of limits}
\label{sec:analysis}

No excess above SM background is observed.
We set an upper bound on the $\tstarpair$ production cross section using the asymptotic modified frequentist $\text{CL}_{\text{s}}$ criterion~\cite{Junk:1999kv, Read:2002hq, asymptotic_cls:Cowan:2010js, ATLAS:2011tau}.
The null hypothesis likelihood function is taken from the background component of the signal+background fit described in Section \ref{sec:background}.
For the uncertainties described in Section~\ref{sec:uncertainties}, a joint template is used, where the nominal template is linearly interpolated to the templates generated with the relevant parameters shifted by ${\pm}1$ standard deviation.
Each of the interpolation variables is taken as a nuisance parameter with a standard Gaussian prior.

\begin{figure}
\centering
\includegraphics[width=\cmsFigWidthLarge]{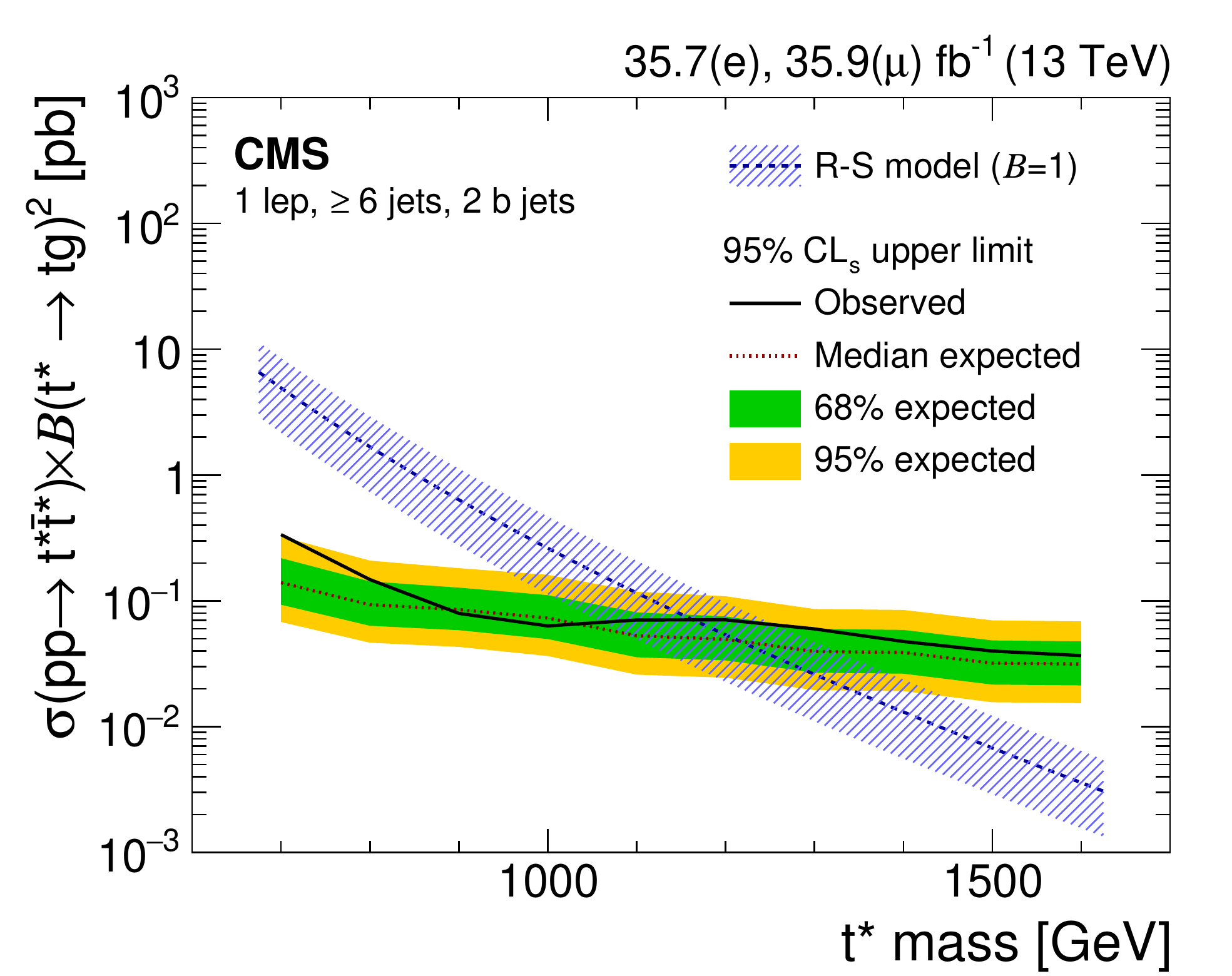}
\caption{
The expected and observed 95\% confidence level upper limits for the product of the production cross section of $\tstarpair$ 
and the square of the branching fraction, as a function of the $\tstar$ mass, for the combined lepton+jets analysis.  
The theoretical production cross section assuming a 100\% $\tstar\to\PQt\Pg$ branching fraction is shown along with its uncertainties, described in Section~\ref{sec:uncertainties}.
}
\label{fig:limit}
\end{figure}

The fit is performed separately in the muon and electron channels, and the results of both are used to obtain combined limits. 
Figure~ \ref{fig:limit} shows the observed and expected upper limits at 95\% confidence level for the product of the $\tstarpair$ production cross section and the square of the branching fraction, as a function of the $\tstar$  mass.
The lower limit for $ m_{\tstar}$ is given by the value at which the upper limit intersects with the theoretical cross section from Ref. \cite{tstarxsec:Stirling:2011ya}. Both the observed and expected lower limits of $ m_{\tstar}$ for the combined muon and electron data are 1.2\TeV, within uncertainties.

\section{Summary}
\label{sec:summary}

A search has been conducted for pair production of spin-\threehalf excited top quarks $\tstar$ in proton-proton interactions, with each \tstar~decaying exclusively to a standard model top quark and a gluon.
Events that have a single muon or electron and at least six jets, exactly two of which must be identified as originating from a bottom quark, are selected for the analysis. Assuming \tstarpair~production, 
the final-state objects are associated with the \tstar~candidates in each event.
No significant deviations from standard model predictions are observed in the $\PQt$+jet~system, and an upper limit is set at 95\% confidence level on the pair production cross section of \tstarpair, as a function of the \tstar~mass.  Interpreting the results in the framework of a spin-\threehalf \tstar~model, assuming a $100\%$ branching fraction of its decay into a top quark and a gluon, \tstar~masses below 1.2\TeV are excluded.
These are the best limits to date on the mass of spin-\threehalf excited top quarks and the first at 13\TeV.

\begin{acknowledgments}

We congratulate our colleagues in the CERN accelerator departments for the excellent performance of the LHC and thank the technical and administrative staffs at CERN and at other CMS institutes for their contributions to the success of the CMS effort. In addition, we gratefully acknowledge the computing centres and personnel of the Worldwide LHC Computing Grid for delivering so effectively the computing infrastructure essential to our analyses. Finally, we acknowledge the enduring support for the construction and operation of the LHC and the CMS detector provided by the following funding agencies: BMWFW and FWF (Austria); FNRS and FWO (Belgium); CNPq, CAPES, FAPERJ, and FAPESP (Brazil); MES (Bulgaria); CERN; CAS, MoST, and NSFC (China); COLCIENCIAS (Colombia); MSES and CSF (Croatia); RPF (Cyprus); SENESCYT (Ecuador); MoER, ERC IUT, and ERDF (Estonia); Academy of Finland, MEC, and HIP (Finland); CEA and CNRS/IN2P3 (France); BMBF, DFG, and HGF (Germany); GSRT (Greece); OTKA and NIH (Hungary); DAE and DST (India); IPM (Iran); SFI (Ireland); INFN (Italy); MSIP and NRF (Republic of Korea); LAS (Lithuania); MOE and UM (Malaysia); BUAP, CINVESTAV, CONACYT, LNS, SEP, and UASLP-FAI (Mexico); MBIE (New Zealand); PAEC (Pakistan); MSHE and NSC (Poland); FCT (Portugal); JINR (Dubna); MON, RosAtom, RAS, and RFBR (Russia); MESTD (Serbia); SEIDI and CPAN (Spain); Swiss Funding Agencies (Switzerland); MST (Taipei); ThEPCenter, IPST, STAR, and NSTDA (Thailand); TUBITAK and TAEK (Turkey); NASU and SFFR (Ukraine); STFC (United Kingdom); DOE and NSF (USA).

\end{acknowledgments}

\bibliography{auto_generated}

\cleardoublepage \appendix\section{The CMS Collaboration \label{app:collab}}\begin{sloppypar}\hyphenpenalty=5000\widowpenalty=500\clubpenalty=5000\textbf{Yerevan Physics Institute,  Yerevan,  Armenia}\\*[0pt]
A.M.~Sirunyan, A.~Tumasyan
\vskip\cmsinstskip
\textbf{Institut f\"{u}r Hochenergiephysik,  Wien,  Austria}\\*[0pt]
W.~Adam, F.~Ambrogi, E.~Asilar, T.~Bergauer, J.~Brandstetter, E.~Brondolin, M.~Dragicevic, J.~Er\"{o}, A.~Escalante Del Valle, M.~Flechl, M.~Friedl, R.~Fr\"{u}hwirth\cmsAuthorMark{1}, V.M.~Ghete, J.~Grossmann, J.~Hrubec, M.~Jeitler\cmsAuthorMark{1}, A.~K\"{o}nig, N.~Krammer, I.~Kr\"{a}tschmer, D.~Liko, T.~Madlener, I.~Mikulec, E.~Pree, N.~Rad, H.~Rohringer, J.~Schieck\cmsAuthorMark{1}, R.~Sch\"{o}fbeck, M.~Spanring, D.~Spitzbart, W.~Waltenberger, J.~Wittmann, C.-E.~Wulz\cmsAuthorMark{1}, M.~Zarucki
% ...
\vskip\cmsinstskip
% ...
1:~~Also at Vienna University of Technology, Vienna, Austria\\
% ...

\end{sloppypar}
\end{document}